\documentclass[twoside]{ae100prg}
\usepackage{graphicx}
\usepackage[breaklinks]{hyperref}
\usepackage{booktabs}

\begin{document}
\title{Effect of magnetic fields on equatorial circular orbits in Kerr spacetimes}

%% \author{Your Name}
%% \address{Affiliation, Address}
%% \ead{your@email.address}

\author{Ignacio F. Ranea-Sandoval$^{1,2}$ and H\'ector Vucetich$^1$}

\address{$^1$ Grupo de Gravitaci\'on, Astrof\'isica y Cosmolog\'ia, Facultad de Ciencias Astron\'omicas y Geof\'{\i}sicas -
 Universidad Nacional de La Plata. Paseo del Bosque S/N (1900). La Plata, Argentina.}
\address{$^2$ CONICET, Rivadavia 1917, 1033, Buenos Aires, Argentina.}

\email{iranea@fcaglp.unlp.edu.ar}

\begin{abstract}
In this work we analyze the effects of an external magnetic field on charged particles on equatorial circular orbits in a Kerr spacetime, both in the black hole and the naked singularity cases. Understanding these phenomena is of great importance because equatorial circular orbits are a key ingredient of simple accretion disc models. We focus on two important magnetic field configurations: a) a uniform magnetic field aligned with the angular momentum and b) a dipolar magnetic field. We center our attention on the effect of these external fields on the marginally bounded and marginally stable equatorial circular orbits because they give information on  observable quantities that could be useful to determine whether the central object is a black hole or a naked singularity. Using a perturbative approach we are able to give analytic results and compare (in the black hole case) with previous results.
\end{abstract}

\section{Introduction}

Since Penrose proposed what is now known as the  {\em {Cosmic Censorship Conjecture}}, 
  spacetimes with naked singularities have been  the subject of  great debate.
 In \cite{2012CQGra..29i5017D} and references therein the authors  prove the existence of unstable perturbations for the most relevant nakedly singular spacetimes. These and other works clearly favor black holes over naked singularities to model extremely compact objects.

Following an alternative  approach, in \cite{1980BAICz..31..129S} the differences between circular geodesics around a Kerr black hole and a Kerr naked singularity are studied.

\begin{figure}
  \begin{center}
  \includegraphics[width=81mm]{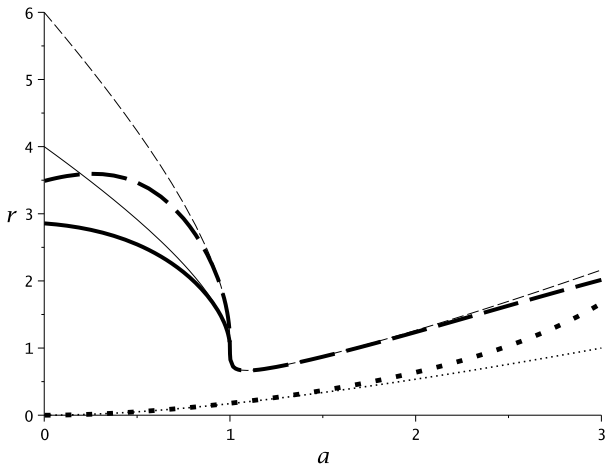}
  \end{center}
  \caption{\label{fig1} Radii of the marginally bounded orbit (solid or dotted lines) and of the innermost stable orbit (dashed lines) as a function of the rotation parameter $a$. Equal line thickness lines correspond to equal values of the parameter $\lambda$ which measures the coupling between the matter's electric charge and the external magnetic field strenght: the thinnest for $\lambda = 0$ and the thickest for $\lambda =0.1$. }
\end{figure}

Accretion processes are usually associated with compact objects and are the main tools astrophysicists possess to study their properties. 
The presence of magnetic fields in accretion discs produce observable phenomena which are worth examining.

The effects of magnetic fields on accretion discs around a rotating black hole where studied, and the changes in the innermost stable orbit and in the marginally bound orbits are analyzed in \cite{1985Prama..25..135I,1978Prama..11..359P,1983JPhA...16.2077W}. 
As we are only able to observe the effects of the presence of a black hole on particles, changes in these particular radii may give observable quantities that could allow us to distinguish between different theoretical models for compact objects. 

In this work we present some of the analytical results of \cite{enprep} in which the change in the position of the inner edge of an accretion disc in a 
Kerr spacetime is studied. We generalize previous results by allowing the rotation parameter $a$ to adopt values larger that one. For sake of space limitation we will not present mathematical details, which are left to the extended work \cite{enprep}.

\begin{figure}
  \begin{center}
  \includegraphics[width=81mm]{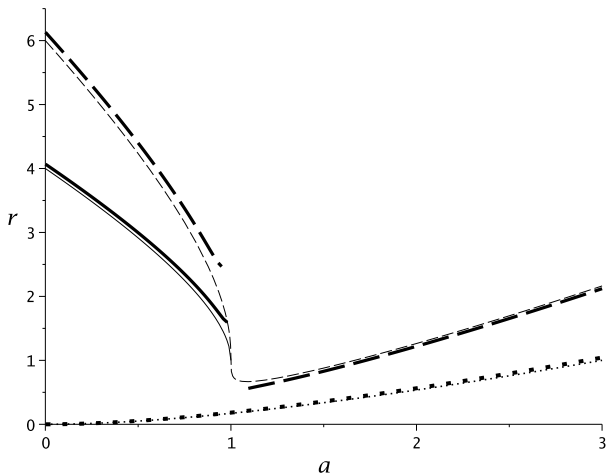}
  \end{center}
  \caption{\label{fig2} Radii of the marginally bounded orbit and of the innermost stable orbit as a function of the rotation parameter $a$. The line styles and thickness correspond to the ones presented in Figure \ref{fig1}.}
\end{figure}

\section{Theoretical Basics}

Kerr spacetime represents the exterior gravitational field of a rotating body and is one of the most important exact solutions to Einstein's
 field equations. Stationary magnetic field configurations in a Kerr background have been studied in detail in \cite{1976GReGr...7..959B,2011PhRvD..83l4046M,1975PhRvD..12.2218P}. The study of the motion of charged particles in the equatorial plane in the presence  fields that preserve the Killing nature of both $\partial_\phi$ and $\partial_t$ is usually done using variations of the arguments in the pioneering works of Carter \cite{1968PhRv..174.1559C} for uncharged particles.

\section{Results}

We use an effective potential  and a perturbative approach that allow us to investigate analytically the cases in which the coupling between the external magnetic field and the effective charge of the particles is small. As a result we obtain analytic expressions for the position of the innermost stable and marginally bounded radii of equatorial circular orbits.

As is explained in \cite{1985Prama..25..135I,1978Prama..11..359P,1983JPhA...16.2077W}, the coupling between the magnetic field strength and the effective charge of particles is measured by a parameter called $\lambda$. Our perturbative approach allows us to study the $\lambda \ll 1$ regime which correspond to the astrophysically relevant case of a (neutral over large scales) simple fluid disc.

We summarize part of our results in Figure \ref{fig1} for the case of a uniform magnetic field, and in Figure \ref{fig2} for the dipolar case. These results are in complete agreement with those obtained for black holes in \cite{ 1985Prama..25..135I,1983JPhA...16.2077W}. 
%The dipolar magnetic field is the first approximation to the more complex and realistic intrinsic magnetic field configuration of astronomical objects.
% This fact makes the study of the dipolar configuration particularly relevant.

\section{Conclusions}

Using a perturbative approach we are able to reproduce analytically previously reported 
 numerical results for the radii of the most relevant circular orbits of charged particles orbiting a Kerr black hole with an external magnetic field,  and extend these results  to the superspinning case. 
The study of the properties of more complex disc models can be an important tool to test observationally the validity of Penrose's Cosmic Censorship Conjecture, and to properly understand the proposed disc-jet relationship.

\section*{References}
\bibliography{biblio01}

\end{document}